# Numerical investigation of the vortex breaker for a dynamic separator using Computational Fluid Dynamics

Rim GUIZANI[1]†, Hatem MHIRI[1], Philippe BOURNOT[2]

[1] *LTTPI, Laboratory of Thermic and Thermodynamic Industrial Processes, National Engineering School of Monastir (ENIM), 5019 Monastir, TUNISIA*

[2] *Aix-Marseille University, CNRS, IUSTI, 13 013 Marseille, FRANCE*

†*Corresponding Author Email: rim.guizani@issatso.u-sousse.tn*

## ABSTRACT

The separation efficiency and pressure drop of the dynamic separator of cement particles can be affected by many factors, like structural type, geometric parameters, and operating characteristics. In this paper, CFD modeling is applied to investigate the fluid flow behavior and the efficiency of the industrial dynamic separator with different heights of the inner cone called the vortex breaker. Simulations are based on the RSM and the DPM models. A CFD comparison of the original design and new designs has been performed. The simulation results showed that the fluid flow inside the industrial air separator is greatly dependent on the height of the vortex breaker. Interesting phenomena were observed by the numerical simulations and the results revealed that an increase in the height of the vortex breaker up to three-quarters of the magnitude of the fine powder outlet duct can improve the performances of particle separation not only by reducing 29% the cut size, and by 40% the bypassing of fine particles but also by increasing 30% the separation sharpness while keeping the pressure drop substantially unchanged.

**Keywords**: CFD simulation; dynamic separator; fish-hook effect; geometry design; optimization

### NOMENCLATURE

| | | | | |
|---|---|---|---|---|
| $Cj$ | Configuration j | | $P$ | Pressure, Pa |
| *CFD* | Computational Fluid Dynamics | | PSD | Particle Size Distributions |
| d | Diameter of the rotor, m | | rpm | revolution per minute |
| $d_{25}, d_{75}$ | Particle diameters, µm | | *RSM* | Reynolds Stress Model |
| $d_{50}$ | Cut size, µm | | *s* | *Sharpness Index* |
| $D$ | Separator diameter, m | | *u* | velocity, m/s |
| DPM | Discrete Phase Model | | x, y, z | axes |
| $H$ | Height of the fine outlet, m | | | |
| $h_j$ | *height of the vortex breaker, m* | | j=1, 2, 3 | |
| *Hj* | *Separator dimensions, m* | | | |
| LES | Large Eddy Simulation | | | Kinematic viscosity, m²s⁻¹ |
| MRF | Multiple Reference Frame | | | |

## 1. INTRODUCTION

A high-efficiency dynamic separator is used to control the final product fineness of the grinding circuit. This equipment was developed mainly for the high demands of the minerals, food, pharmaceutical, and cement industries. In this study, we focused on the dynamic separator used in cement industry plants.

The dynamic call comes from the highly rotating rotor inside the huge separator which is used to improve the centrifugal force of fine cement particles to be separated easily in powder classification processes.

The performance of the rotor separators is evaluated by analyzing the selectivity or the tromp curve which describes the percentage of chance of a given particle diameter coming from the inlet feed of the separator and recovered in the coarse particle outlet as a function of the particle size.

The experimental efficiency curve of an air separator reveals a special profile that happens from recapturing fine particles in the heavier coarse particle outlet. This occurrence is known as the fish-hook effect and affects almost all types of industrial separators.

This phenomenon reduces the performances of the air classifiers hence many researchers focused to reduce it through experimentation and numerical studies.

So far, the most available literature were dealing with the fish-hook effect in hydrocyclones. Ghodrat. M et al. (2014) proved by simulation that using a long convex cone enhanced the performance of a hydrocyclone. Jiang et al. (2020) proposed a new W-shaped structure of a hydrocyclone to achieve a better efficiency curve.

A few studies focused on this phenomenon in rotor separators. For instance, an experimental and numerical study was performed by Huang. Q et al. (2012) to evaluate the flow field characteristics of the turbo air classifier with curved guide blades. Results proved that a little cut-size and a greater classification precision index are produced with positively bowed guide blades. A previous work made by Eswaraiah. C et al. (2012) established that the fish-hook happening depends on the conception of the guide vanes and the fluid flow patterning.

Another study was made by Liu. R et al. (2015) where authors proved experimentally and by CFD simulation that the classification performance of a turbo air classifier is increased with the new conception of the inclined guide vanes.

Over the last few years, CFD simulations were applied to analyze the complicated flow inside the air separators aiming to find a solution to this problem as reported in many studies. Such as, Huang Q. et al. (2012) used numerical simulations to study the turbo air classifier. Guo L. et al. (2013) also proved by CFD simulations that the proposed bottom structure of the rotor cage of the turbo air classifier could produce a narrow PSD of cement powder.

Just recently, Misiulia et al. (2020) showed by CFD simulations that adding a scroll and radial bend have significant effects on the separation zone of a high-efficiency separator. Zhang et al. (2020) investigated a greater removal efficiency cyclone by injecting vapor into fine particles. Sun et al. (2022) have studied the performances of a horizontal classifier with three rotor cages used in silicone powder. Good prediction accuracy was found using the RSM and the DPM models compared to experimental gas-solid flow field results.

Until now many studies have focused on the development of a new design of cyclone separators. Li et al. (2022) proved by simulation that improving the cylinder roof structure reduces the top short-circuit flow and makes a more efficient vortex of the cyclone separator.

Zhang. L et al. (2022) performed a numerical simulation of a cyclone separator with pulverized powder coal. Calculation results showed that the proposed lower exhaust could enhance the separation efficiency. Another numerical and experimental study made by Zhang, R et al. (2022) proved that when an apex cone is installed at the dust outlet, the back-mixing of particles will be decreased but the separation efficiency will be enhanced.

However, none of the researchers has focused on the study of the geometry of the vortex breaker cone in the inner cylinder of the rotor separator.

In this paper, we report a numerical investigation of the effects of the inner vortex breaker size on the separation performances and the flow field characteristics of the industrial separator. CFD simulations are used to analyze the turbulent flow field and to predict the separation characteristics of five different configurations with a rotating rotor. Performances, including the selectivity curve, the cut size, the sharpness index, and the by-pass are then compared. The complexity of particle separation flow behavior inside the dynamic separator still possess challenges in making a new economical and more efficient design.

## 2. NUMERICAL METHODS

### 2.1 Geometry structures

Figure 1. Illustrates the schematic view of the initial rotor separator and a sliced plane showing the inner vortex breaker. The geometry has been simulated with actual sizes that can be found in Table 1 as reported by Guizani R et al. (2014). The most important element of this study is the vortex breaker cone. It was placed in the central part of the separator near the four outlets of the fine particles. Five different values of h/H were tested as illustrated in table 1. We used the Cartesian coordinate system is in the simulation study. The center of the upward coarse particle outlet was settled to be the origin of the coordinate system as it is shown in Figure 2. C2 is the referenced configuration installed in the cement plant.

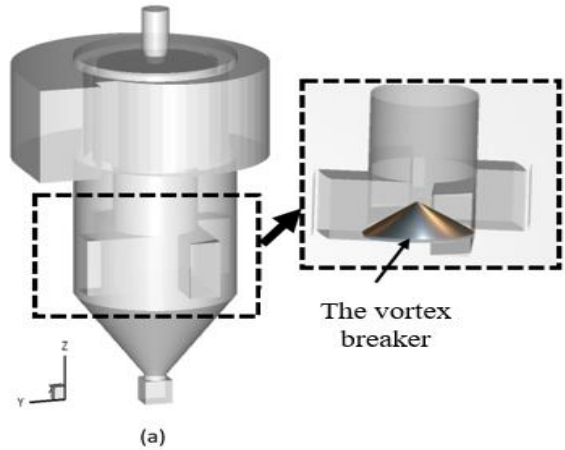

**Fig. 1. A vertical view of the simulated dynamic separator.**

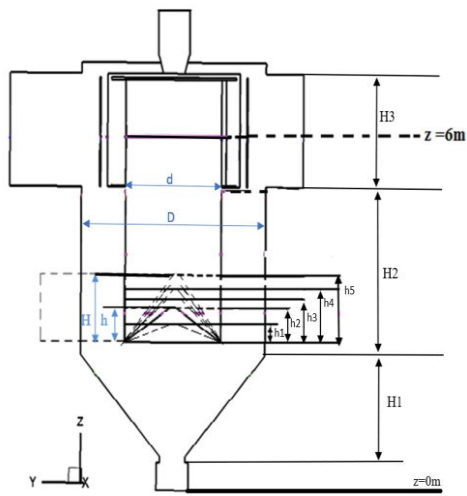

**Fig. 2. Schematic diagram of the different tested configurations of the dynamic separator**

**Table1. Summary of the tested configurations' dimensions and ratios.**

| Dimensions (mm) | Values |
|---|---|
| Separator diameter, D | 4040 |
| Internal cylinder diameter, d | 2100 |
| Height of the fine outlet, H | 1420 |
| Height of the cone, H1 | 1720 |
| Height of the cylinder body, H2 | 2880 |
| Inlet volute height, H3 | 1936 |
| Configuration1 C1, h1/H | 0.25 |
| Configuration 2 C2, h2/H | 0.5 |
| Configuration 3 C3, h3/H | 0.66 |
| Configuration 4 C4, h4/H | 0.75 |
| Configuration 5 C5, h5/H | 1 |

## 2.2 Model description

To study the complex fluid flow behavior in the different configurations of the dynamic separator, we used a hybrid grid with a combination of tetrahedral and hexahedral cells. A special refinement was adopted in the critical zones like the rotary rotor cage, the twelve guide blades, and the gas inlet volute.

Figure 3 shows the meshes and the boundary conditions in the dynamic separator. The air inlet was set as a velocity inlet of 18,25 ms$^{-1}$ which was calculated from the industrial data. All the outlets are set to pressure outlets, which were in ambient condition. All walls were set with no-slip boundary conditions. The speed of the rotor reaches 111 rpm. The density of the cement powder is 1190 kg/m3.

The authors have already developed a process model to validate the efficiency curve and the pressure drop between the inlet and the outlet of the separator. The predicted results were validated with the experimental data in a previous work reported by Guizani R et al. (2014). Hence, the Reynolds Stress Model (RSM) of Ansys Fluent 18.0 is used under the viscous model because of the incompressible turbulent flow field in the industrial device. In recent studies, The Large Eddy Simulation (LES) model is proven to be more accurate in recirculating flows as published by Rajmistry. S et al. (2017) and Brar. L et al. (2020). However, this model requires an important amount of computer resources and highly accurate spatial and temporal discretization.

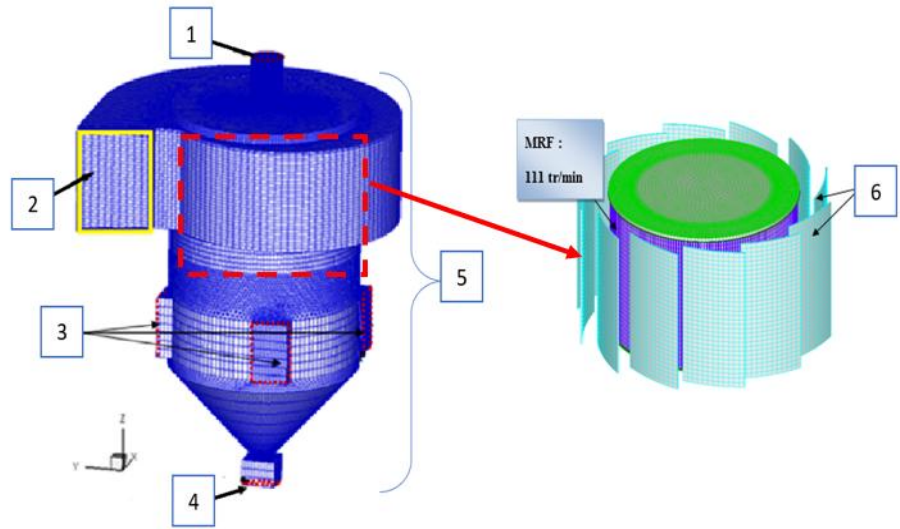

1-Particle's feed inlet 2- Air inlet 3- Fine particles outlets 4- Coarse particles outlet 5- Wall 6-Guide blades.
**Fig. 3. Meshes and boundary conditions of the dynamic separator.**

It was also proved that the RSM model provides the goodness of fit of the results compared with the experiment as it was reported by Huang. Q et al. (2012), and Guo. L et al. (2013). The MRF technique was used to simulate the rotational motion of the rotor cage. The calculations were carried out in an unsteady state. In this study, the convergence of the RSM model was only possible if we choose a very small time step size. We used a 5 $e^{-4}$-time step size in all calculations. The steady state of the average static pressure calculated at the air inlet surface was attained after 2 seconds. Therefore, the overall calculation time was preferred to be 2 seconds.

We applied the coupled method of the discrete phase with the continuous gas phase.

Based on the complex flow field in the rotor separator, we divided the calculation process into two steps. In the primary step, only the air phase was considered. The gas phase characteristics were calculated using the RSM model. The distribution of the velocity and the pressure profiles were calculated. The next step was carried out by introducing the different sizes of the solid cement powder as the discrete phase using the DPM model. In this step, the gas-solid flow and the selectivity curve were investigated to analyze and compare the separation performances of the different configurations of the industrial separator. The descriptions of the RSM, the MRF, and the DPM models can be found in references Guizani. R et al. (2014) and (2017), so it is not repeated here.

Turbulent flows like in gas separators are greatly influenced by the existence of the walls. Modeling the turbulence near the walls requires a very fine mesh with adapted meshing strategies. The y+ adaption scheme is used in this study. The approach is to calculate y+ for boundary cells on the specified viscous wall regions, define the value between the minimum and the maximum allowable, and mark and adapt the appropriate cells. The y+ is commonly used to define the law of the wall. In fact, the wall y+ is a non-dimensional number like the local Reynolds number. It determines whether the influences in the wall-adjacent cells are laminar or turbulent, hence indicating the part of the turbulent boundary layer that they resolve. Y+ is given by Eq (1).

$$y^+ = u_T \, y/\nu$$

(1)

Where y is the distance to the nearest wall, $u_T$ is the friction velocity, and $\nu$ is the local kinematic viscosity of the fluid used. In this study, wall grid adaption is employed for y+ between 30 and 60.

The mesh independence test was made for the C2 separator design to ensure that the results are appropriate. Three numbers of cells were tested: 757,366, 932,521, and 1,204,290. The convergence of mesh 3 requires a longer time and more computer resources although the calculated pressure drop difference between mesh 2 and mesh 3 can be disregarded as shown in Figure 4. Therefore, we select mesh 2 with 991,320 cells for the overall simulation of the industrial separator.

In the other configurations, we have nearly the same number of cells.

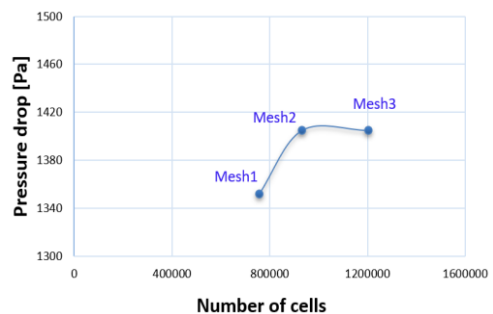

**Fig. 4. Grid independence test of the dynamic separator.**

## 3. SIMULATION RESULTS AND ANALYSIS

In this section, Numerical simulation has been used to study the effects of the variation of the height of the vortex breaker on the flow structure and performances of the dynamic separator.

### 3.1 Effects on the internal flow structure

Figure 5 shows the time-averaged pressure distributions in the yz plane at x=0m and the xy plane at z=6m for the same air inlet flow rate inside the different configurations of the considered dynamic separator. The obtained results show that the variation in the height of the "vortex breaker" highly affects the internal structure of the flow in the dynamic separator. This difference is especially noticeable in the axial central zone where a strong depression appears for all the configurations. This depression zone is occurred because of the moving motion of the rotor cage in addition to the vortices localized on the axis of the flow. In the same figure, we noticed that this axis of the flow takes a specific structure for each height of the vortex breaker. Furthermore, there is a marked increase in the static pressure near the walls of the cyclonic body of the separator since all the mass was ejected against the wall. Therefore, the deep pressure gradients are observed in the radial direction of each separator. These gradients decrease gradually towards the outlet of coarse particles to the bottom of the equipment. The minimum pressure is localized on the axis of the separator where the flow is highly turbulent, and the Reynolds Stress Closure model is the most appropriate to predict the pressure accurately Mokni. I et al. (2015) and Sun et al. (2017).

Figure 6 clearly describes the streamlines of the airflow on the yz plane at x=0m for the various configurations studied. Results show that the height of the vortex breaker greatly affects the streamlines of the highly swirled fluid flow localized inside the rotating cage and up to the outlet of fine particles. Another important result is the occurrence of the recirculation flows. These recirculation zones are intensive due to the creation of the vortex and the high Reynolds Number. They were stabilized on the axis of each separator, and they took the form of recirculation bubbles. This physical phenomenon is called the "vortex breakdown". It was clearly observed near the geometric axis of the rotary separator. The location and shape of these recirculation balls highly depend on the size of the vortex breaker. Indeed, the more the height of the vortex breaker is increased, the more the recirculation bubbles move away from the central zone of the rotating cage. From this analysis, we can suggest that the third and the fourth configurations could be more efficient than the reference configuration C2 that is used by the cement industry. We noticed that these recirculation flows deteriorate the performances of the dynamic separator. Therefore, we will analyze deeply in the following sections, the effect of these recirculation flows in the classification performances.

As the velocity distribution should be responsible for the separation phenomenon, in this type of equipment, it is very important to analyze the effect of the vortex breaker size on the components of the velocity.

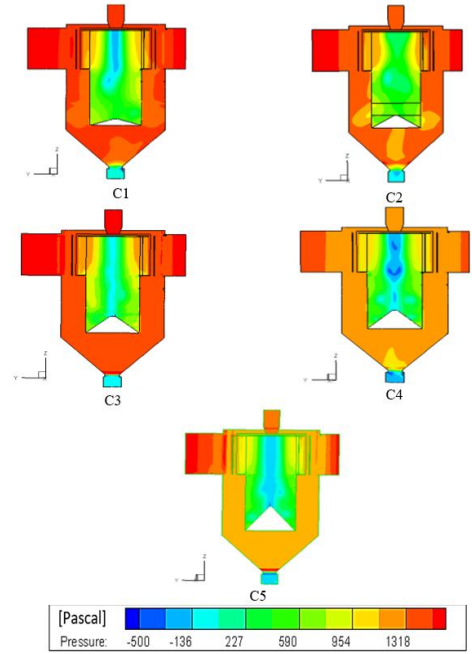

**Fig. 5. Radial pressure contours of the different configurations.**

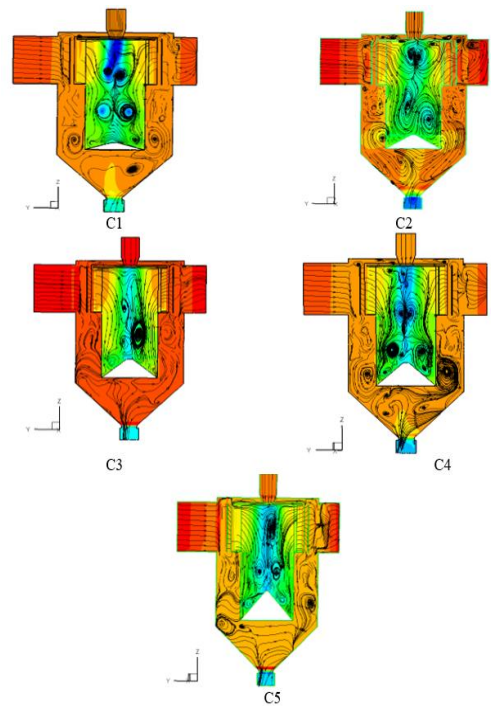

**Fig. 6. The predicted streamlines and recirculation flows inside the tested configurations**

### 3.2 Effects on the velocity distribution

In cyclone separators, tangential velocity, and radial velocity were always used to determine the stability of the flow field. It was also proved that increasing the tangential velocity will enhance the powder moving speed in the annular region and the centrifugal force field as reported by Liu et al. (2015).

Figure 7 describes the predicted time-averaged tangential velocity contours for the different configurations. The velocity values are represented by different colors. Indeed, the red color represents the maximum values while the blue one represents the negative value, and the green color is close to zero. The air inlet velocity is 18.25 m·s⁻¹. Results showed that the tangential velocity depends clearly on the vortex breaker size. Indeed, the rotation of the rotor increases the tangential velocity produced by the tangential inlet velocity component that induces the centrifugal force. The latter is the most useful for the classification of cement particles from the gas-solid mixture. As the separating air enters the rotating cage, it begins to accelerate tangentially. Therefore, the tangential velocity reached its maximum value up to 1.5 to 2 times the inlet velocity value for all the studied configurations. It takes these optimal values near the rotor cage and more precisely on a cylinder radius that is half of the inner cylinder radius as shown in Figure 7. The increasing of the tangential velocity creates the rotational or the vortex with great intensity near the rotor cage and decreases gradually towards the lower part of the separator due to the friction effect of the walls. Consequently, this weakening will be compensated by an increase in the axial velocity component to hold the momentum balance.

The upward axial velocity in the separation zone was proved experimentally to be detrimental to classification in rotor separators Guo et al. (2007), Sharf et al. (2014) and Jiang et al. (2019). While this velocity component in the inner cylinder will lead to a change in the airflow velocity direction and increase the instability of the flow field Liu et al. (2011).

Consequently, a larger upward axial velocity makes it easy for some particles to return to the upper part near the rotor cage to cause excessive particle loading and enhance particle-particle collision and agglomeration. Therefore, the regression of the classification performance was mainly affected by this phenomenon as reported by Guo et al. (2007).

The predicted time-averaged axial velocity contours of the five different cases studied in this paper are shown in Figure 8. Results showed that the maximum values of the axial velocity are localized in the central axis of the separator for all the configurations. Furthermore, a downward axial velocity appeared near the inner cylinder wall (negative values). These instabilities of the axial velocity led to the creation of the recirculation zones as seen in Figure 6. We also note that these secondary flows induce the particles already selected to be selected again which forms the fish-hook effect in the selectivity curves. Despite this description of the flow and the analysis of the velocity field for the different configurations, the choice of an optimal configuration remains difficult and requires a comparison of the selectivity performances which are the pressure drop and the selectivity curve.

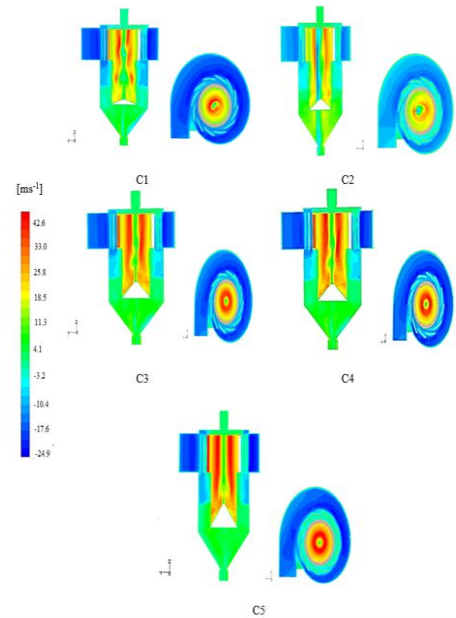

**Fig. 7. Tangential velocity contours of the different configurations**

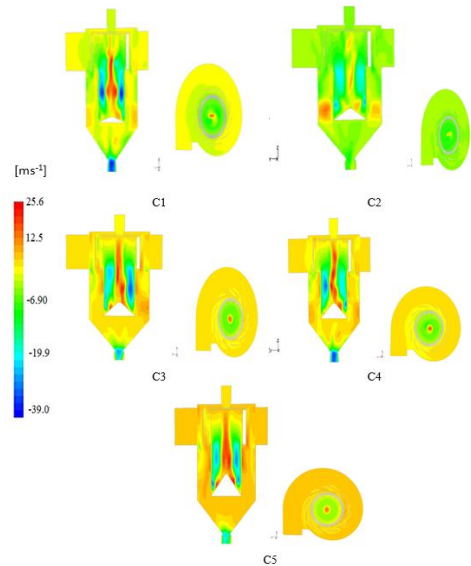

**Fig.8. Axial velocity contours of the different configurations**

### 3.3 Effects on the pressure drop

The pressure drop is a very important characteristic for evaluating the efficiency of the separation since it is directly related to operating costs. It is calculated as the difference in the static pressure between the inlet and the outlet of the separator. The improvement of the separation efficiency could be well evaluated only if the pressure drop remains constant or reduced. Figure 9 shows the evolution of the pressure drops across the five different studied separators. Results showed that the vortex breaker size affects the pressure drop. It was clear that the pressure drop increased with the increase in the height of the vortex breaker. The configuration choice depends on the optimal efficiency as well as the pressure drop minimum values. Therefore, we must refer to the selectivity curve and a comparison of the associated parameters such as the cut size diameter, the sharpness index, and the bypass to make the right choice.

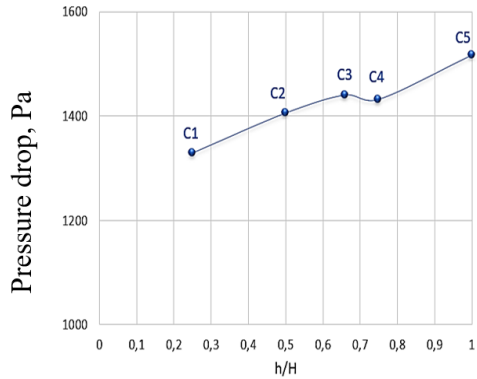

**Fig. 9. The effect of the vortex breaker on the pressure drop.**

### 3.4 Effects on the selectivity curve

For the same cumulative particle size distribution of the mass inlet separator for the different cases studied in this paper, the selectivity curves are calculated. The distribution of the feed inlet which is calculated by the sieving method is provided by the industrial cement plant as shown in Figure 10. Effectively, particle simulations are made with different numbers of cement particles varying from 2300 to 20,000. By keeping identical particle size distributions for all the configurations, we deduced that the predicted efficiency results are not influenced by the number of particles injected as reported by Guizani. R et al. (2014). After validation of the efficiency curve for the studied separator, the calculated efficiency curves of the five cases are shown in Fig 11(a).

The total separation efficiency refers to the recovery of all solid particles in the coarse particle outlet compared to those in the feed inlet. In general, the industrial selectivity curve decreases monotonically with the particle size and asymptotes to a given value called the by-pass. It was the lowest point of the selectivity curve which forms the fish-hook effect. This phenomenon was seen for the C1 and C2 configurations of the studied separator.

According to this study, it was proved that the fish-hook effect was markedly reduced from the third configuration. For the rest, the selectivity curve becomes an increasing function of the particle size.

The separation performances can be deduced from the selectivity curves as it is illustrated in Figure 11(b). Specifically, the cut size $d_{50}$, sharpness index s, and the by-pass $\tau$. These characteristics are picked up from the selectivity curve, calculated for the different designs, and then compared in Figure 12.

The cut size $d_{50}$ is defined as the critical particle size corresponding to a separation efficiency of 50% on the efficiency curve Jiang et al. (2019) that is particles of this diameter have an equal chance of entering either the coarse particle outlet or the fine particle outlet. The sharpness index s is calculated as follows by Ivan et al. (1986).

$$s = \frac{d_{25}}{d_{75}} \qquad (2)$$

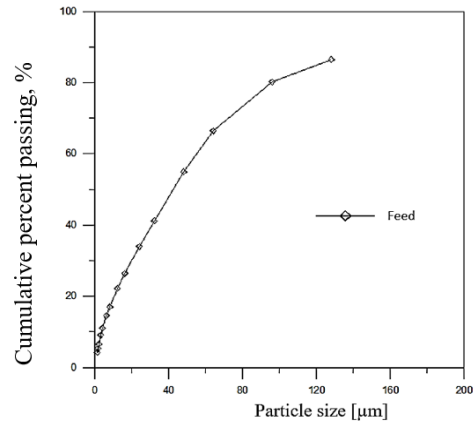

**Fig. 10. The industrial distribution of the feed particles**

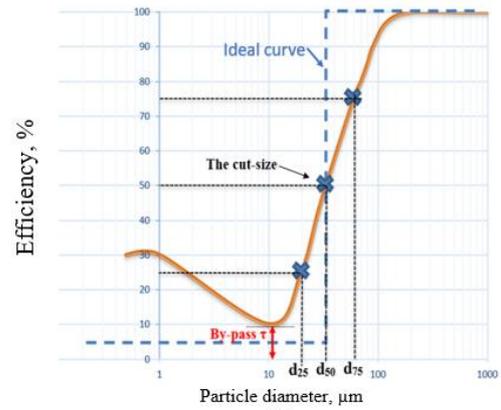

(a)

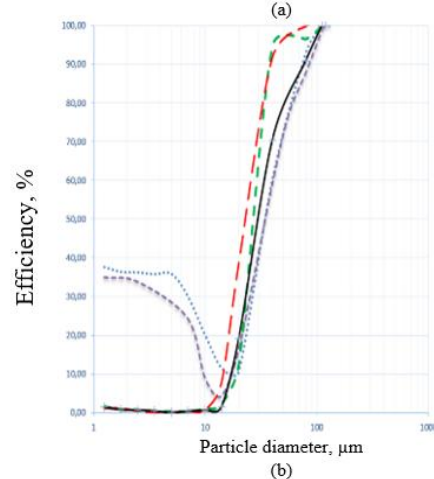

(b)

**Fig. 11(a) Comparison of the numerical selectivity curves of the different studied**

**configurations (b) The selectivity curve and separation performances of the air separator**

When the sharpness s is closer to 1, the classification process will be closer to the ideal separation curve. The by-pass is a measure of the percentage of the deviation of fine particles in the coarse outlet due to secondary flows in the inner cylinder as it is shown in Figure 11(b). It can also be used as a measure of the quality of the separation. Indeed, better efficiency will be achieved when the by-pass of fine particles is the lower as possible.

Figure 12 (a) shows that the cut size highly depends on the height of the vortex breaker. It first decreases and then increases. The cut size reaches its minimum value for the C4 configuration, which is 29.4 % lower than that of reference C2.

Based on the analysis of the sharpness index in Figure 12 (b), it is found that the increase of the cone size can improve by 30% the separation sharpness. On one hand, the two configurations C3 and C4 have two sharpness indexes relatively greater than that of the conventional design separator. On the other hand, the analysis of the by-pass factor in Figure 12 (c) shows that the by-pass decreases gradually by increasing the height of the vortex breaker cone. Configuration C4 reduces the bypass by 40 % compared with the referenced configuration. From all these data it is becoming increasingly clear that the choice of the fourth configuration could conduct a better separation performance.

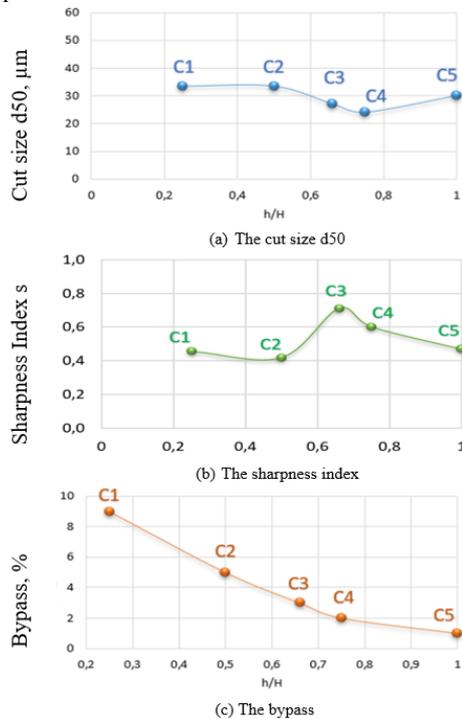

(a) The cut size d50

(b) The sharpness index

(c) The bypass

**Fig. 12. Comparison of separation performances.**

## 4. CONCLUSIONS

In the present investigation, the CFD method is applied to numerically simulate and analyze the separation performances of the rotor separator with five different heights of the vortex breaker. The major findings are stated below:

(1) For the same boundary and operating conditions, the height variation of the vortex breaker cone highly affects the internal structure of the turbulent flow in the industrial air separator.

(2) Interesting phenomena such as the recirculating flows and the vortex breakdown have been observed by numerical simulation of the internal flow.

(3) It has also been proved that an increase in the size of the vortex breaker up to three-quarters of the height of the fine outlet tube improves the selectivity efficiency by reducing the by-passing of the fine particles with the coarse ones. This result contributes to eliminating the fish-hook effect in the efficiency curve so that it reduces the cut-off size.

(4) It is also proved that making the vortex breaker longer increases by 30% the sharpness of the separation and reduced the cut size by 29% while keeping the pressure drop substantially unchanged.

These results are of great interest since they could possibly improve separation performances at a lower cost compared with experimental optimization. Without forgetting that geometry optimization of the rotor separator would provide a new thought for the design of other separation devices.

## REFERENCES

Brar, Lakhbir Singh, and J. J. Derksen (2020). Revealing the details of vortex core precession in cyclones by means of large-eddy simulation. *Chemical Engineering Research and Design* 159, 339-352.

Eswaraiah, C., S. I. Angadi, and B. K. (2012). Mishra. Mechanism of particle separation and analysis of fish-hook phenomenon in a circulating air classifier. *Powder Technology* 218, 57-63.

Ghodrat, M., Kuang, S. B., Yu, A. B., Vince, A., Barnett, G.D., & Barnett, P. J. (2014). Numerical analysis of hydrocyclones with different conical section designs. *Minerals Engineering* 62, 74-84.

Guizani, R., Mokni, I., Mhiri, H., & Bournot, P. (2014). CFD modeling and analysis of the fish-hook effect on the rotor separator's efficiency. *Powder Technology* 264, 149-157.

Guizani, R., Mhiri, H., and Bournot, P. (2017). Effects of the geometry of fine powder outlet on pressure drop and separation performances for dynamic separators. *Powder Technology* 314, 599-607.

Guo, L., Liu, J., Liu, S., & Wang, J. (2007). Velocity measurements and flow field characteristic analyses in a turbo air classifier. *Powder Technology* 178.1, 10-16.

Huang, Qiang, Jiaxiang Liu, and Yuan Yu. (2012). Turbo air classifier guide vane improvement

and inner flow field numerical simulation. *Powder Technology* 226, 10-15.

Klumpar Ivan V., Currier Fred N., Ring Terry A. (1986). Air classifiers. Chemical Engineering 3, 77-92.

Jiang, L., Liu, P., Yang, X., Zhang, Y., Wang, H., & Xu, C. (2019). Numerical analysis of flow field and separation characteristics in hydrocyclones with adjustable apex. Powder Technology 356, 941-956.

Jiang, L., Liu, P., Yang, X., Zhang, Y., & Wang, H. (2020). Designing W-shaped apex for improving the separation efficiency of a full-column hydrocyclone. Separation Science and Technology 55(9), 1724-1740. Designing W-shaped.

Li, W., Huang, Z., Li, G., & Ye, C. (2022). Effects of different cylinder roof structures on the vortex of cyclone separators. Separation and Purification Technology 296, 121370.

Liu, G-F., Yuan Yu, and J-X. Liu. (2011). Effect of volute with horizontal plates on flow fields in turbo air classifiers. Chemical Engineering 39(7), 69-73.

Liu, Rongrong, Jiaxiang Liu, and Yuan Yu (2015). Effects of axial inclined guide vanes on a turbo air classifier. Powder Technology 280, 1-9.

Misiulia, D., Antonyuk, S., Andersson, A. G., &Lundström, T. S. (2020). High-efficiency industrial cyclone separator A CFD study. Powder Technology 364, 943-953.

Mokni, I., Dhaouadi, H., Bournot, P., & Mhiri, H. (2015) Numerical investigation of the effect of the cylindrical height on separation performances of uniflow hydrocyclone. Chemical Engineering Science 122, 500-513.

Rajmistry, S., Ganguli, S., Chandra, P., & Chatterjee, P. K. (2017). Numerical Analysis of Gas-Solid Behavior in a Cyclone Separator for Circulating Fluidized Bed System. Journal of Applied Fluid Mechanics 10, 1167-1176.

Sharf, Abdusalam M., Hosen A. Jawan, and Fthi A. Almabsout (2014). The influence of the tangential velocity of inner rotating wall on axial velocity profile of flow through vertical annular pipe with rotating inner surface. EPJ Web of Conferences 67. EDP Sciences.

Sun, Z., Liu, C., Yang, G., & Chen, L. (2022). Orthogonal vortices characteristic, performance evaluation and classification mechanism of a horizontal classifier with three rotor cages. Powder Technology 404, 117438.

Sun, Z., Sun, G., Liu, J., & Yang, X. (2017). CFD simulation and optimization of the flow field in horizontal turbo air classifiers. Advanced Powder Technology 28 (6), 1474-1485.

Zhang, L., Chen, Y., Zhao, B., Dang, M., & Yao, Y. (2022). Numerical simulation on structure optimization of escape-pipe of cyclone separator with downward outlet. Powder Technology, 117588.

Zhang, R., Yang, J., Han, S., Hao, X., & Guan, G. (2022). Improving advantages and reducing risks in increasing cyclone height via an apex cone to grasp vortex end. Chinese Journal of Chemical Engineering.

Zhang, Y., Yu, G., Jin, R., Chen, X., Dong, K., Jiang, Y., & Wang, B. (2020). Investigation into water vapor and flue gas temperatures on the separation capability of a novel cyclone separator. Powder Technology 361, 171-178.